\documentclass[10pt]{article}

\usepackage{amsmath}
\usepackage{amssymb}

\usepackage{graphicx}
\usepackage{multirow}
\usepackage{cite}

\usepackage{color}

\usepackage{setspace} 
\usepackage[labelfont=bf,labelsep=period,justification=raggedright]{caption}

\makeatletter
\renewcommand{\@biblabel}[1]{\quad#1.}
\makeatother

\date{}

\begin{document}

% Title must be 150 characters or less
\begin{flushleft}
{\Large
\textbf{Ashes 2013 $-$ A network theory analysis of Cricket strategies}
}
% Insert Author names, affiliations and corresponding author email.
\\
\bf {Satyam Mukherjee}$^{1}$
\\
{1} Kellogg School of Management, Northwestern University, Evanston, IL, United States of America
\\
{1} Northwestern Institute on Complex Systems (NICO), Northwestern University, Evanston, IL, United States of America
\\

%$\ast$ E-mail: amaral@institute.edu
\end{flushleft}

% Please keep the abstract between 250 and 300 words
\section*{Abstract}
We demonstrate in this paper the use of tools of complex network theory to describe the strategy of Australia and England in the recently concluded Ashes $2013$ Test series. 
Using partnership data made available by cricinfo during the Ashes $2013$ Test series, we generate batting partnership network (BPN) for each team, in which nodes correspond to batsmen and links represent runs scored in partnerships between batsmen.  The resulting network display a visual summary of the pattern of run-scoring by each team, which helps us in identifying potential weakness in a batting order. We use different centrality scores to  quantify the performance, relative importance and effect of removing a player from the team. We observe that England is an extremely well connected team, in which lower order batsmen consistently contributed significantly to the team score. Contrary to this Australia showed dependence on their top order batsmen. 
\section*{Cricket as a complex network}

\begin{figure*}
\begin{center}
\includegraphics[width=14.5cm]{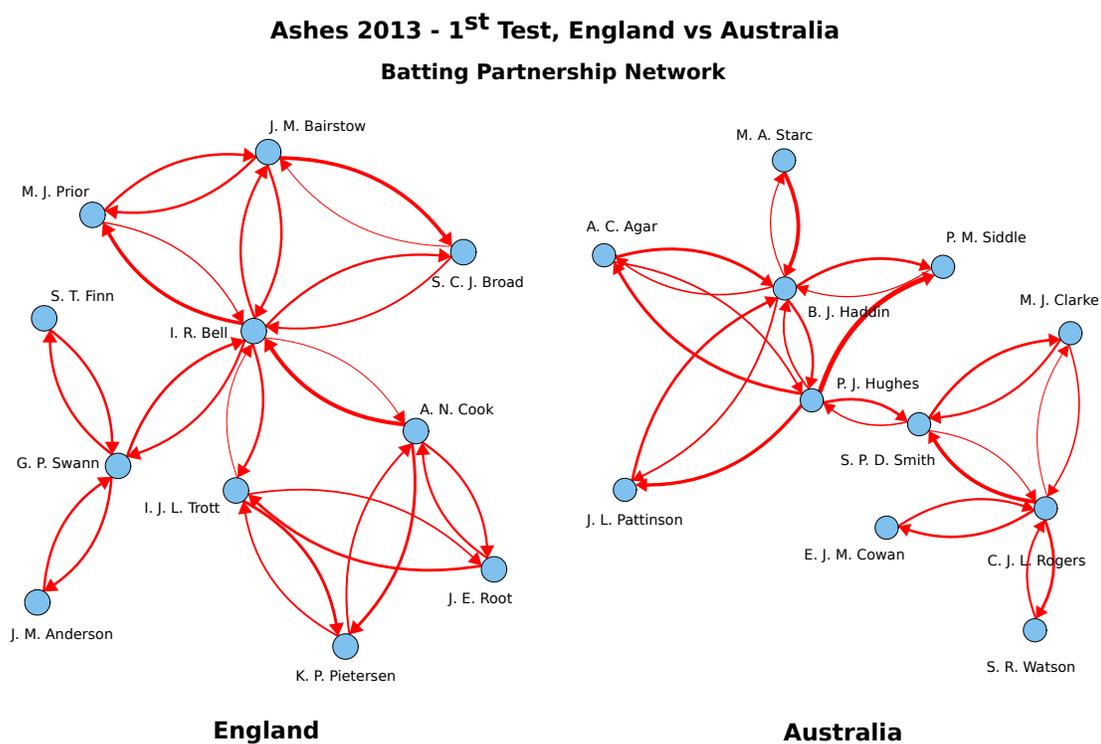}
\caption{ \label{fig:directed} Batting partnership network for the first Ashes Test involving England and Australia. Visually we observe the difference in scoring pattern for the two teams. For Australia, the dependence on BJ Haddin, PJ Hughes and SPD Smith is evident. The situation is different for England, where IR Bell formed regular partnership with the top order players like AN Cook and also lower order players like GP Swann. England displayed this team work throughout the Ashes 2013. While for Australia the scoring was mostly dependent on the top order players. }
\end{center}
\end{figure*}

The tools of social network analysis have previously been applied to sports. Such as a network approach was developed to quantify the performance of individual players in soccer \cite{duch10} and football \cite{pena}. Network analysis tools have been applied to football \cite{girvan02}, Brazilian soccer players \cite{onody04}, Asian Go players \cite{xinping}. Successful and un-successful performance in water polo have been quantified using a network-based approach \cite{mendes2011}. Head-to-head matchups between Major League Baseball pitchers and batters was studied as a bipartite network \cite{saavedra09}. More recently a network-based approach was developed to rank US college football teams \cite{newman2005}, tennis players \cite{radicchi11} and cricket teams and captains \cite{mukherjee2012}. 

The central goal of network analysis is to capture the interactions of individuals within a team. Teams are defined as groups of individuals collaborating with each other with a common goal of winning a game. Within teams, every team member co-ordinate across different roles and the subsequently influence the success of a team. In the game of Cricket, two teams compete with each other. Although the success or failure of a team depends on the combined effort of the team members, the performance or interactive role enacted by individuals in the team is an area of interest for ICC officials and fans alike. We apply network analysis to capture the importance of individuals in the team. The game of Cricket is based on a series of interactions between batsmen when they bat in partnership or when a batsman is facing a bowler. Thus a connected network among batsmen arises from these interactions. 

In cricket two batsmen always play in partnership, although only one is on strike at any time. The partnership of two batsmen comes to an end when one of them is dismissed or at the end of an innings. In cricket there are $11$ players in each team. The batting pair who start the innings is referred to as opening-pair. For example two opening batsmen $A$ and $B$ start the innings for their team. In network terminology, this can be visualized as a network with two nodes $A$ and $B$, the link representing the partnership between the two players. Now, if batsman $A$ is dismissed by a bowler, then a new batsman $C$ arrives to form a new partnership with batsman $B$. This batsman $C$ thus bats at position number three. Batsmen who bat at positions three and four are called top-order batsman. Those batsmen who bat at positions  five to seven are called middle-order batsmen. Finally the batsmen batting at position number eight to eleven are referred to as tail-enders. 
Thus a new node $C$ gets linked with node $B$. In this way one can generate an entire network of batting-partnership till the end of an innings. The innings come to an end when all $11$ players are dismissed or when the duration of play comes to an end. The score of a team is the sum of all the runs scored during a batting partnership. In our work we considered the network of batting partnership for Australia and England during the Ashes $2013$ Test series.

\subsection*{Performance Index}
We generate weighted and directed networks of batting partnership for all teams, where the weight of a link is equal to the fraction of runs scored by a batsman to the total runs scored in a partnership with another batsman. Thus if two batsmen $A$ and $B$ score $n$ runs between them where the individual contributions are $n_A$ and $n_B$, then a directed link of weight $\frac{n_{A}}{n}$ from $B$ to $A$. In Figure~\ref{fig:directed} we show an example of weighted and directed batting partnership network for two teams - Australia and England. We quantify the batting performance of individual players within a team by  analyzing the centrality scores - in-strength, PageRank score, betweenness centrality and closeness centrality.

For the weighted network the in-strength $s_{i}^{in}$ is defined as 
\begin{equation}
s_{i}^{in} = \displaystyle\sum_{j \ne i} \omega_{ji}
\end{equation}
where $\omega_{ji}$ is given by the weight of the directed link.

We quantify the importance or `popularity' of a player with the use of a complex network approach and evaluating the PageRank score. Mathematically, the process is described by the system of coupled equations
\begin{equation}
    p_i =  \left(1-q\right) \sum_j \, p_j \, \frac{{\omega}_{ij}}{s_j^{\textrm{out}}}
+ \frac{q}{N} + \frac{1-q}{N} \sum_j \, \delta \left(s_j^{\textrm{out}}\right) \;\; ,
\label{eq:pg}
\end{equation}
where ${\omega}_{ij}$ is the weight of a link and $s_{j}^{out}$ = $\Sigma_{i} {\omega}_{ij}$ is the out-strength of a link. $p_i$ is the PageRank score assigned to team $i$ and represents the fraction of the overall ``influence'' sitting in the steady state of the diffusion process on vertex $i$ (\cite{radicchi11}).  $q \in \left[0,1\right]$ is a control parameter that  awards a `free' popularity to each player and $N$ is the total number of players in the network. 
The term $ \left(1-q\right) \, \sum_j \, p_j \, \frac{{\omega}_{ij}}{s_j^{\textrm{out}}}$  represents the portion of the score received by node $i$ in the diffusion process obeying the hypothesis that nodes  redistribute their entire credit  to neighboring nodes. The term $\frac{q}{N}$ stands for a uniform redistribution of credit among all nodes. The term $\frac{1-q}{N} \, \sum_j \, p_j \, \delta\left(s_j^{\textrm{out}}\right)$ serves as a correction in the case of the existence nodes with null out-degree, which otherwise would behave as sinks in the diffusion process.  It is to be noted that the PageRank score of a player depends on the scores of all other players and needs to be evaluated at the same time. To implement the PageRank algorithm in the directed and weighted network, we start with a uniform probability density equal to $\frac{1}{N}$ at each node of the network. Next we iterate through  Eq.~(\ref{eq:pg}) and obtain a steady-state set of PageRank scores for each node of the network. Finally, the values of the PageRank score are sorted to determine the rank of each player. According to tradition, we use a uniform value of $q=0.15$. This choice of $q$ ensures a higher value of PageRank scores \cite{radicchi11}.

Another performance index is betweenness centrality, which measures the extent to which a node lies on a path to other nodes. In cricketing terms, betweenness centrality measures how the run scoring by  a player during a batting partnership depends on another player. Batsmen with high betweenness centrality are crucial for the team for scoring runs without losing his wicket. These batsmen are important because their dismissal has a huge impact on the structure of the network.  So a single player with a high betweenness centrality is also a weakness, since the entire team is vulnerable to the loss of his wicket. In an ideal case, every team would seek a combination of players where betweenness scores are uniformly distributed among players. Similarly the opponent team would seek to remove the player with higher betweenness centrality. 
Closeness centrality measures how easy it is to reach a given node in the network \cite{wasserman,peay}. In cricketing terms, it measures how well connected a player is in the team. Batsmen with high closeness allow the option for changing the batting order depending on the nature of the pitch or match situation. 

 In  Table~\ref{tableS3} we compare the performance of players for different teams. 
 In the first Test, according to PageRank, in-strength, betweenness and closeness measures {\it IR Bell} is the most successful batsman for England. The betweenness centrality of {\it IR Bell} is the highest among all the players. This is also supported by the fact that in the second innings of the first Test at Trent Bridge, he scored $109$ when England were $121$ for $3$. He added $138$ crucial runs for the seventh wicket with Broad, and helped England post a target of $311$. According to the centrality score {\it IR Bell} was the most deserving candidate for the Player of the match award.   
In the second Test, England's {\it JE Root}, with the highest in-strength, betweenness and closeness emerges as the most successful of all batsmen. Even his PageRank score is third best among all the batsmen, with Australia's {\it UT Khawaja} and {\it MJ Clarke} occupying the first and second spot. Interestingly {\it JE Root} was declared the Player of the match. In the third Test at Old Trafford, Australia dominated the match. The dominance is reflected in the performance of {\it MJ Clarke} ( highest PageRank, in-strength, betweenness and closeness among all batsmen) and quite deservedly received the player of the match award. In the fourth Test we observe that for Australia the centrality scores are not evenly distributed. For example {\it CJL Rogers} has the highest PageRank, in-strength, betweenness and closeness, while the other players have much lower centrality scores. The pattern of run scoring is dependent heavily on the top order batsmen. On the contrary for England we observe that a lower order batsman like {\it TT Bresnan} has the highest PageRank, indicating strong contribution from the lower order batsmen. Similarly the betweenness scores are much more uniformly distributed $-$ a sign of well balanced batting unit. Lastly, in the final Test, we again observe a strong dependence on top order batsmen like {\it SR Watson} and {\it SPD Smith} in the Australian team. Both {\it SPD Smith} and {\it SR Watson} are the two most central players. The remaining batsmen have either small or zero betweenness scores. Looking into the scores of England, we observe that  middle order batsman {\it IR Bell} is the most central player, {\it MJ Prior} has the highest closeness, top order batsman {\it IJL Trott} has the highest in-strength and finally lower order batsman {\it GP Swann} has the highest PageRank. Thus there is no predominant role in the English batting line-up, indicating a well connected team as compared to the Australian batting line-up. We also observe that in terms of centrality scores, England's IR Bell performed consistently in most of the Test matches and also shared the Player of the series with {\it RJ Harris} of Australia. 
Thus we see that tools of social network analysis is able to capture the consensus opinion of cricket experts. The details of the procedure is available online \cite{satyamacs} and has been accepted for publication in Advances in Complex Systems.

\begin{table}[ht!]
\centering
\caption{{Performance of batsmen for England and Australia for Ashes $2013$ } The top five performers are ranked according to their PageRank score and their performance is compared with In-strength, betweenness centrality and closeness centrality.}
\begin{tabular}{cccccc}

\hline
\multirow{2}{*}{}
 {Country} & {Players} & {PageRank} & {In-strength}& {Betweenness} & {Closeness}\\ 
\hline 

\multirow{3}{*}{{Australia}}	
	& \multirow{6}{*}{}	
	BJ Haddin & 0.2327 & 2.51818 & 0.4666  & 0.4761\\
 & CJL Rogers & 0.1296  & 1.4099 &  0.4555 & 0.4545\\
Test $1$ & PJ Hughes & 0.1087 & 1.1771  & 0.5555 & 0.5882\\
 &SPD Smith & 0.1045 & 1.7468  & 0.5333 & 0.5555 \\
 & PM Siddle & 0.0967 & 1.55  & 0.0  & 0.3333 \\
\hline 

\multirow{3}{*}{{England}}	
	& \multirow{6}{*}{}	
	IR Bell & 0.1924 & 2.6740 & 0.7555 & 0.7142\\
 & GP Swann &0.1208 & 1.4666 &  0.3777 & 0.5263 \\
Test $1$ & IJL Trott & 0.1043 & 1.4240  & 0.1666 & 0.5263\\
& JM Bairstow & 0.1038 & 1.1257  & 0.0111 & 0.4761 \\
 & AN Cook & 0.0840 & 1.0473  & 0.1666  & 0.5263 \\
\hline 
\hline
\multirow{3}{*}{{Australia}}	
	& \multirow{6}{*}{}	
	UT Khawaja & 0.2188 & 1.5897 & 0.3444  & 0.4166\\
 & MJ Clarke  & 0.2088  & 2.2086 &  0.3666 & 0.2777\\
Test $2$& SPD Smith & 0.1355 & 1.1176  & 0.0 & 0.1923\\
 &CJL Rogers & 0.0846 & 0.5549  & 0.1444 & 0.3125\\
 &BJ Haddin & 0.0754 & 0.8179  & 0.3333  & 0.4545 \\

\hline 

\multirow{3}{*}{{England}} 
	& \multirow{6}{*}{}
	JE Root & 0.1882 & 2.4721 & 0.6111  & 0.625\\
 & SCJ Broad & 0.1306  & 0.9583 &  0.2000 & 0.3333\\
Test $2$& TT Bresnan & 0.0946 & 1.2727  & 0.4666 & 0.5555\\
 & KP Pietersen & 0.0908 & 1.625  & 0.0 & 0.4 \\
 & JM Anderson & 0.0862 & 1.0  & 0.3555  & 0.4347 \\
\hline 
\hline
\multirow{3}{*}{{Australia}}	
	& \multirow{6}{*}{}	
	MJ Clarke & 0.2513 & 2.9611 & 0.7142  & 0.8\\
 &CJL Rogers & 0.1590  & 2.5731 &  0.1071 & 0.6153\\
Test $3$& BJ Haddin & 0.1235 & 1.4841  & 0.25 & 0.5714\\
 & SR Watson & 0.1113 & 1.6  & 0.1785 & 0.5714 \\
 & MA Starc & 0.1042 & 1.2304  & 0.0  & 0.5333 \\
\hline 

\multirow{3}{*}{{England}}	
	& \multirow{6}{*}{}	
	MJ Prior & 0.1787 & 1.5252 & 0.5333  & 0.5555\\
 & KP Pietersen & 0.1757  & 2.6446 &  0.6777 & 0.6666\\
Test $3$& AN Cook & 0.1537 & 2.2147  & 0.2777 & 0.5263\\
 & JE Root & 0.0970 & 1.3035  & 0.1 & 0.5263 \\
 & GP Swann & 0.0749 & 0.7333  & 0.0  & 0.3703 \\

\hline 
\hline
\multirow{3}{*}{{Australia}}	
	& \multirow{6}{*}{}	
	CJL Rogers& 0.2195 & 3.1408 & 0.7  & 0.7142\\
 &BJ Haddin & 0.1523  & 1.9342 &  0.2666 & 0.4545\\
Test $4$& SR Watson & 0.1423 & 1.7771  & 0.0 & 0.4545\\
 & DA Warner & 0.0849 & 1.0184  & 0.2777 & 0.5 \\
 & MJ Clarke & 0.0797 & 1.6383  & 0.2777  & 0.5 \\
\hline 

\multirow{3}{*}{{England}}	
	& \multirow{6}{*}{}	
 TT Bresnan & 0.2146 & 2.0091 & 0.5333  & 0.5263\\
 & IR Bell & 0.1422  & 2.6809 &  0.5333 & 0.5555\\
Test $4$& JM Bairstow & 0.1291 & 1.2647  & 0.2 & 0.4761\\
 & KP Pietersen & 0.0906 & 1.1675  & 0.3111 & 0.4761 \\
 & AN Cook & 0.0902 & 1.2164  & 0.2  & 0.5 \\
\hline 
\hline
\multirow{3}{*}{{Australia}}	
	& \multirow{6}{*}{}	
	SPD Smith & 0.2155 & 3.6472 & 0.5222  & 0.7363\\
 & SR Watson & 0.1734  & 2.9791 &  0.3333 & 0.6231\\
Test $5$& MJ Clarke & 0.1576 & 2.1258  & 0.1 & 0.6231\\
 & JP Faulkner & 0.1313 & 2.7275  & 0.0 & 0.54 \\
 & RJ Harris & 0.0706 & 1.2333  & 0.0  & 0.45 \\
\hline 

\multirow{3}{*}{{England}}	
	& \multirow{6}{*}{}	
 GP Swann & 0.1777 & 1.6097 & 0.3778  & 0.4166\\
 & MJ Prior & 0.1367  & 1.5 &  0.6 & 0.5263\\
Test $5$& IJL Trott & 0.1310 & 2.0510  & 0.3555 & 0.4761\\
 & JM Anderson & 0.1091 & 0.8  & 0.0 & 0.4761 \\
 & IR Bell & 0.0945 & 1.1711  & 0.6444  & 0.3030 \\
\hline

\end{tabular}
\label{tableS3}
\end{table}

%\clearpage
\section*{Conclusion}
To summarize, we investigated the structural properties of batsmen partnership network (BPN) in the Ashes 2013 Test series. Our study reveals that  network analysis is able to examine individual level network properties among players in cricket. 
The batting partnership networks not only provides a visual summary of proceedings of matches for various teams, they are also used to analyze the importance or popularity of a player in the team. We identify the pattern of play for Australia and England and potential weakness in batting line-up. Identifying the `central' player in a batting line up is always crucial for the home team as well as the opponent team. We observe that Australia depended more on the top batsmen, while England was involved in team-game. 

There are some additional features which could be applied in our analysis. The networks in our study are static and we assumed all the batsmen are equally athletic in the field. One could add an ``athletic index" as an attribute to each batsman. Also adding the fielders as additional nodes in the networks could provide us with a true picture of the difficulty faced by a batsman while scoring.
In real life many networks display community structure : subsets of nodes having dense node-node connections, but between which few links exist. Identifying community structure in real world networks have could help us to understand and exploit these networks more effectively. Potentially our study leaves a wide range of open questions which could stimulate further research in other team sports as well.

\section*{Acknowledgements}

The author thanks the cricinfo website for public availability of data.

\end{document}